\documentclass[sigconf,bookmarksnumbered,unicode,nonacm,bookmarks=false]{acmart}
\AtBeginDocument{%
  \providecommand\BibTeX{{%
    \normalfont B\kern-0.5em{\scshape i\kern-0.25em b}\kern-0.8em\TeX}}}

\usepackage{alltt}
\usepackage{booktabs}
\usepackage{graphicx}
\usepackage{framed}
\usepackage{float}
\usepackage[abbreviations,british]{foreign}  
\usepackage{listings}
\usepackage{multicol}
\usepackage{multirow}
\usepackage{pifont} 
\usepackage{subcaption}
\usepackage{soul} 
\usepackage{tikz}
\lstdefinelanguage{Gherkin}
{
  morekeywords={
    And,
    Scenario,
    Given,
    When,
    Then
  },
  sensitive=true
}

\newfloat{lstfloat}{htbp}{lop}
\floatname{lstfloat}{Listing}

\newcommand\numUnitTests{481}

\newcommand\numTestCases{53}

\title{Did You Remember To Test Your Tokens?}

\author{Danielle Gonzalez}
\affiliation{%
  \institution{Rochester Institute of Technology}
  \streetaddress{1 Lomb Memorial Drive}
  \city{Rochester, NY}
  \country{USA}}
\email{dng2551@rit.edu}

\author{Michael Rath}
\affiliation{%
  \institution{DLR Institute of Data Science}
  \institution{Technical University Ilmenau}
  \city{Jena / Ilmenau}
  \country{Germany}
}

\author{Mehdi Mirakhorli}
\affiliation{%
  \institution{Rochester Institute of Technology}
  \streetaddress{1 Lomb Memorial Drive}
  \city{Rochester, NY}
  \country{USA}}

\renewcommand{\shortauthors}{Gonzalez et al.}

\date{January 2019}
\begin{document}

\begin{abstract}
Authentication is a critical security feature for confirming the identity of a system's users, typically implemented with help from frameworks like Spring Security. It is a complex feature which should be robustly tested at all stages of development. Unit testing is an effective technique for fine-grained verification of feature behaviors that is not widely-used to test authentication. Part of the problem is that resources to help developers unit test security features are limited. Most security testing guides recommend test cases in a ``black box" or penetration testing perspective. These resources are not easily applicable to developers writing new unit tests, or who want a security-focused perspective on coverage.

In this paper, we address these issues by applying a grounded theory-based approach to identify common (unit) test cases for token authentication through analysis of 481 JUnit tests exercising Spring Security-based authentication implementations from 53 open source Java projects.
The outcome of this study is a developer-friendly unit testing guide organized as a catalog of 53 test cases for token authentication, representing unique combinations of 17 scenarios, 40 conditions, and 30 expected outcomes learned from the data set in our analysis. We supplement the test guide with common test smells to avoid. To verify the accuracy and usefulness of our testing guide, we sought feedback from selected developers, some of whom authored unit tests in our dataset.
\end{abstract}

\maketitle
%
\section{Introduction}
Authentication is a critical security feature for controlling who has access to the potentially sensitive data or operations of a software system~\cite{fernandez2013security,stallings2012computer,buildingsecuresoftware}. When authentication is included in a system's \textit{security design}, it is typically implemented using language-specific frameworks (\eg{} Spring Security, Apache Shiro) rather than built `from scratch'~\cite{KazmanFrameworks}. However, even with a framework, implementing authentication is complicated because executing typical functionality (\eg{} submitting a login form) requires multiple components (\eg{} objects) to interact~\cite[Ch.\ 5, p.\ 6]{fernandez2013security}. Further, there is no `single design' for authentication because several mechanisms (\eg{} password, PIN), types (\eg{} token, session), and protocols (\eg{} SAML~\cite{rfc7522}, LDAP~\cite{rfc4513}) exist, and systems can be designed to handle more than one configuration. This flexibility is reflected in authentication APIs, and adds more complexity to implementation.

With complexity comes risk, especially for security features; \textit{broken authentication} is second in OWASP's \textit{Top 10 Most Critical Web Application Security Risks}~\cite{van2017owasp}. Hence, the importance of testing complex \& critical security features is clear. The most common security testing techniques, penetration and functional testing~\cite{potter2004software}, are ``black-box'': performed on running instances of a system with no knowledge of the source code. However, there is an established ``white-box'' (code available) technique that can be applied much sooner, before a prototype is even ready: \textit{unit testing}.


Unfortunately, unit tests are often overlooked for security~\cite{DEVRIES200711}; traditional definitions~\cite{5733835} imply that unit tests have limited scope. However, a recent study of JUnit tests by Trautsch \etal{}~\cite{Trautsch2020} found no difference between the defect-detecting capabilities of unit and \textit{integration} tests, traditionally defined as tests of interactions between modules~\cite{5733835}. This is supported by the Google Test Blog, which considers \textit{unit} testing module interactions (`behaviors') a best practice~\cite{GoogleTestBlog}, and other groups advocating more focus on automated software testing for security implementations~\cite{vanDeursen:2015:BPO:2791301.2793039,de2006security, DEVRIES200711}.

Security testing is motivated by risk models, attack patterns, and use/misuse cases~\cite{wysopal2006art,addressingissues,buildingsecuresoftware}. Respected security testing resources, such as the OWASP testing guide~\cite{meucci2013owasp}, suggest test cases in this context from the ``black box'' testing perspective. This unfortunately limits their applicability for developers writing \textit{unit} tests. In practice, unit test planning decisions are made arbitrarily by developers based on their own experience \& education~\cite{smith2008survey}. To make testing more objective, decisions on how \textit{much} to test are often based on metrics such as coverage~\cite{smith2008survey,gonzalez2017large,garousi2010replicated,athanasiou2014test}. Some security APIs have packages for unit testing, but documentation focuses on mocking and test setup~\cite{springtest,shirotest,jaastest} without suggesting test cases.

Currently, there is no empirically-grounded work that addresses this issue by identifying and describing common authentication test cases from a unit testing perspective: combinations of scenarios, conditions, and expected outcomes. Such a resource could be used to guide practitioners towards incorporating security testing earlier in the development life cycle.

In this paper, we address this with \textbf{six novel contributions}: \ding{182} we apply classical grounded theory~\cite{glaser1967discovery} to perform an \textit{inductive analysis} of unit tests for authentication from open source projects. We then \textit{characterize} the authentication scenarios developers commonly test, based on real test data. During data collection, we scoped this investigation to \textit{token} authentication to maintain a distinction between tests for authentication and session management security features.
Our findings were curated into a \ding{183} taxonomy of \numTestCases{} unique test cases, represented as \textit{flow diagrams} and \textit{unit testing test guide}. In addition, the guide is complemented by a \ding{184} set of common test smells found in authentication tests to avoid.
We conduct \ding{185} interviews with two developers who authored tests in our dataset and sought feedback on the guide from 2 other selected practitioners. These took place at the latest stage of our study to verify that the empirical findings are useful in practice. Finally, we provide an online appendix which includes \ding{186} the test guide and \ding{187} a \textit{replication package} containing our unit test dataset, grounded theory artifacts, and supplementary reference data: ~\url{https://doi.org/10.5281/zenodo.3722631}

%
\section{Methodology}
~\label{sec:methods}

A \textit{grounded theory} (GT)~\cite{glaser1967discovery} is a systematic inductive method for conducting qualitative research when \textit{little is known about a phenomenon of interest}. This methodology is driven by \textit{emergence of concepts}~\cite{glaser1992basics,GroundedTheoryInSE}: progressive identification and integration of concepts from data that lead to discoveries directly supported by empirical evidence and grounded in the data~\cite{GroundedTheoryInSE}. Since we do not know in advance the nature of the test cases (except a high-level knowledge that they are related to token authentication), our goal is to allow the data to drive our process of discovering classes of authentication features that need to be tested (inductive reasoning) and their corresponding test cases rather than formulating hypotheses at the beginning of the analysis process (deductive reasoning). In this study we implemented the seven core activities of ``classic'' grounded theory: \textit{identification of topic of interest}, \textit{theoretical sampling}, \textit{data coding} (through \textit{open}, \textit{selective} and \textit{theoretical coding}), \textit{constant comparative analysis}, \textit{memo writing}, \textit{memo sorting} and \textit{write up \& literature review}~\cite{glaser1992basics,GroundedTheoryInSE}.
%
%
In addition, we interviewed practitioners to verify and validate the developer-focused testing guide we developed based on the hypotheses drawn from this study.

\subsection{Phenomena of Interest}
GT is suitable for investigating broad questions like ``\textit{what is going on here?}''~\cite{GroundedTheoryInSE}. Therefore, when following this methodology researchers are advised not to form specific research questions and instead define an \textit{area of interest} (\ie{} the phenomena under observation). For this study, our target phenomena is \textbf{unit testing of authentication implementations}. We focus on authentication because it is a common but complex security feature that receives limited attention in existing unit testing guidelines.

\subsubsection{Definitions\nopunct}
The following are key terms related to our phenomena of interest, with the definitions applied in this study:

\noindent\textbf{Unit Test}: An implementation of a test case, constructed using a common syntax/framework such as JUnit.

\noindent\textbf{Test Case}: A unique combination of a \textit{context}, \textit{action}, \textit{condition}, and \textit{expected outcome}, represented in natural language

\noindent\textbf{Context}: Pre-conditions and test setup \eg{}, object initialization

\noindent\textbf{Action}: Manipulations of the subject-under-test, \eg{} a method call

\noindent\textbf{Condition}: A state or event that influences the action, such as a specific (\eg{} null) parameter value passed to a method call

\noindent\textbf{Expected Outcome}: The desired state of the subject-under-test \textit{after} the action is performed, \eg{} specific method call return value

\noindent\textbf{Scenario}: A single phrase representation of a unique combination of \textit{context} and \textit{action}

\noindent\textbf{Feature}: A label for a set of related test cases

\subsection{Data Collection: Mining Repositories}
\label{sec:mining}
To study our phenomena of interest, we needed access to \textit{authentication unit tests} from real projects. Therefore, we targeted data sources that were freely accessible to us. In this context, we focused on open source projects that implemented \textit{authentication} and have corresponding unit tests.

 To facilitate the use of an automated approach to mine a large and rich dataset for our analysis, we set the following narrowing criteria for language, authentication framework, and testing framework: the projects must be implemented in \textbf{Java}, use the \textbf{Spring Security} framework for authentication, and unit tests must be written using the \textbf{JUnit} framework. Java was selected because it is a popular choice for back-end development of web applications~\cite{jetbrains/developer-ecosystem:2019}.
 The Spring Security framework is a very popular framework used to implement authentication in Java, and JUnit is the mainstream testing framework for Java projects~\cite{gonzalez2017large,jetbrains/developer-ecosystem:2019}.

\noindent\textbf{Data Summary}
The final data set used for analysis consisted of 481 test methods from 125 unique test files, mined from 53 projects. These tests were collected using the following procedure:

\subsubsection{Data Source\nopunct}
We selected GitHub as a repository source. GHTorrent~\cite{Gousi13} and the REAPER tool~\cite{munaiah2017curating} were used to automatically identify, clone, and filter 100,000 randomly selected Java projects from GitHub. The data set was filtered for empty, duplicate, and inactive (no commits or recent updates) repositories using metadata from GitHub API:\@ size, commits, and date of last push. The test ratio metric from REAPER was used to discard projects with no JUnit tests, leaving 33,090 candidate projects. An automated script was developed to parse these projects and verify use of Spring Security, the target authentication framework.

\subsubsection{Detecting Use of Spring Security\nopunct}
A package-search approach was used to identify projects that used Spring Security to implement authentication. To curate a list of relevant packages, we reviewed several forms of documentation provided by Spring: an overview of the framework's architecture~\cite{spring-architecture}, the comprehensive reference document~\cite{spring-ref}, and the API documentation~\cite{spring-api}. First, we identified objects \textit{required} to implement authentication such as an \texttt{Authentication Provider}. These were used as a starting point to find the root packages providing each component in the API.\@ Next, we reviewed \textit{all} root packages in the API documentation to identify packages for flexible components such as the numerous authentication mechanisms (\eg{} LDAP, SAML) Spring Security can integrate with. We excluded packages related to \textit{sessions} to prevent expanding the phenomena under study to session management, which we consider a unique security feature. Our final list of 18 root packages with descriptions has been included in the supplementary materials for this paper:~\url{https://doi.org/10.5281/zenodo.3722631}

\subsubsection{Identifying Relevant JUnit Test \ul{Files}\nopunct}\label{sec:identify-relevant-junit-test-files}
Included in REAPER's repository analysis was a list of JUnit test files for each project. Our script used these lists and our authentication package list to automatically search each test file for imports of one or more of these packages. We addedan additional rule that test files importing the \textit{Security Context Holder} or \textit{User Details} packages should also import at least one more from the list, because these packages are also used to implement authorization. We randomly selected 50 resulting test files to ensure there were no false positives. This search identified 229 test files from 150 projects.

\subsubsection{Duplicate Removal\nopunct}
A major component GitHub's contribution workflow is \textit{forking}, or making a copy, of a repository. Forks in our data set were identified from the GHTorrent metadata. If forks were discovered, the original repository was kept and all forks were discarded. Forks containing unique and relevant tests would have been kept, but this situation did not arise.

We also discovered that several projects in the dataset were built using the JHipster development platform~\cite{JHipster}. This platform helps developers build highly customizable web applications with a Spring-based server side. JHipster also provides a test suite for the code it generates, that includes relevant authentication tests. In our dataset, we found 81 projects that \textit{only} included the provided test file without adding additional related tests to this file or others (as detected by our tools). There was no reason to include duplicates of the same test method in our analysis, so we excluded these projects but included the original file with the authentication test cases from the JHipster repository.

\subsubsection{Identifying Relevant JUnit Test \ul{Methods}\nopunct}
Next, we built another tool that utilized the JavaParser~\cite{javaparser} Symbol Solver to identify \textit{individual test methods} within these files that used objects from the imported packages. This AST parser uses the test file and the source project to identify the package of origin for an object. Our tool was configured to record all test methods that had at least one object originating from one of the 18 root packages on our list, with the additional rule for the \textit{Security Context Holder} and \textit{User Details} packages from test file selection.
To verify the automated identification (see Section~\ref{sec:identify-relevant-junit-test-files}), we performed a manual validation of 50 randomly selected results, and discarded test methods that did not \textit{use} packages from our candidate set.
Our final dataset contained \numUnitTests{} unique test methods.

\subsection{Open Coding \& Memoing}
\begin{figure*}
\centering
\vspace{-10pt}
\includegraphics[clip]{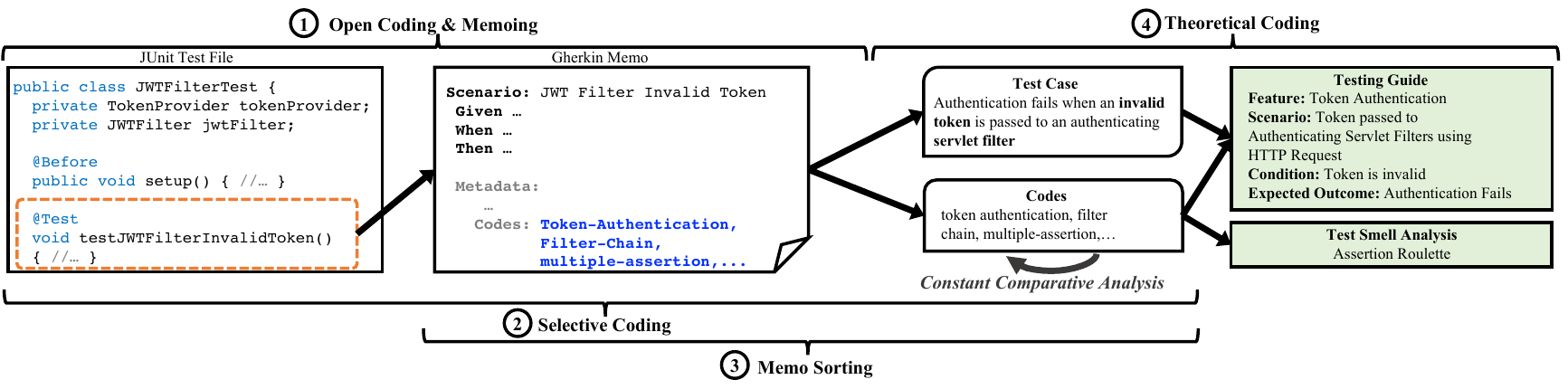}
\caption{Schematic of the applied four step coding process.}
\label{fig:coding-process}
\vspace{-10pt}
\end{figure*}

Two authors independently performed \textit{open coding}~\cite{GroundedTheoryInSE,glaser1992basics} by annotating the 481 test methods in our data set with organically-derived \textbf{codes} reflecting \textit{concepts and themes} present in the test (\cf{} \ding{172} in Figure~\ref{fig:coding-process}).
Authors reviewed each test method and its setup \& helper methods. Accurately identifying \textit{themes} and \textit{concepts} requires thorough understanding of each test method. Prior work in test summarization~\cite{zhang2016} has shown that test comprehension can be achieved by identifying a test's \textit{context}: precondition(s) \& setup; \textit{action}: manipulation(s) of the subject-under-test; and \textit{expected outcome}: desired state and result(s) after performing the action.

We captured these concepts in \textbf{memos} formatted as \emph{Gherkin scenarios}~\cite{gherkinOnline}.
Gherkin is used in behavior-driven development to capture scenarios in a non-technical \& readable format. Test components are represented using three keywords: \emph{Given} (context), \emph{When} (action), \& \emph{Then} (expected outcome). The \textit{And} keyword is used when a components spans multiple statements in a test.
When an existing memo fully represented a new test method, no new memo was created to avoid duplicate memos.
However, the test was still annotated with codes.
We attached the codes as well as more data in a custom \emph{Metadata} section in each scenario (\cf{} bottom block in Listing~\ref{listing:example-memo}), which is not part of the official Gherkin syntax.

Listing~\ref{listing:example-test} shows a unit test \texttt{testJWTFilterInvalidtoken()} that verifies whether a request containing an invalid JSON Web token (JWT) is filtered and no authentication occurs.
Also shown are setup and helper methods from the test class that provide additional context.
Listing~\ref{listing:example-memo} shows the test's corresponding memo in the Gherkin syntax.
The \emph{Token-Authentication} and \emph{Filter-Chain} codes represent core authentication concepts in the test. All memos are included in the supplementary materials.
%
%
\begin{lstfloat}
\begin{lstlisting}[language=Java,
morekeywords={@Test,@Before},
basicstyle=\tiny\ttfamily,
captionpos=b,
escapeinside={(*@}{@*)},
keywordstyle=\color{blue},
numberstyle=\tiny\color{red},
stringstyle=\color{magenta},
commentstyle=\color{gray},
frame=single,
breaklines=true,
label=listing:example-test,
caption=An example JUnit test from the data set and it's corresponding setup \& helper methods.]
public class JWTFilterTest {
  private TokenProvider tokenProvider;
  private JWTFilter jwtFilter;

  @Before
  public void setup() {
    JHipsterProperties jHipsterProperties = new JHipsterProperties();
    tokenProvider = new TokenProvider(jHipsterProperties);
    ReflectionTestUtils.setField(tokenProvider, "secretKey", "test secret");
    ReflectionTestUtils.setField(tokenProvider, "tokenValidityInMilliseconds",
      60000);
    jwtFilter = new JWTFilter(tokenProvider);
    SecurityContextHolder.getContext().setAuthentication(null);
  }

  @Test
  public void testJWTFilterInvalidToken() throws Exception {
    String jwt = "wrong_jwt";
    MockHttpServletRequest request = new MockHttpServletRequest();
    request.addHeader(JWTConfigurer.AUTHORIZATION_HEADER, "Bearer " + jwt);
    request.setRequestURI("/api/test");
    MockHttpServletResponse response = new MockHttpServletResponse();
    MockFilterChain filterChain = new MockFilterChain();

    jwtFilter.doFilter(request, response, filterChain);

    assertThat(response.getStatus()).isEqualTo(HttpStatus.OK.value());
    assertThat(SecurityContextHolder.getContext().getAuthentication()).isNull();
  }
  // ..
}
\end{lstlisting}
\end{lstfloat}

\vspace{-20pt}
%
\begin{lstfloat}
\begin{lstlisting}[language=Gherkin,
basicstyle=\tiny\ttfamily,
captionpos=b,
escapeinside={(*@}{@*)},
% keywordstyle=\color{blue},
frame=single,
breaklines=true,
label=listing:example-memo,
caption=Sample memo \& codes for the test in Listing~\ref{listing:example-test}.]
Scenario: JWT Filter Invalid Token
  Given A new JHipster Properties Object is initialized
  And A Token Provider is initialized with the JHipster Properties Object
  And The Token Provider's secretKey attribute is initialized
  And The Token Provider's tokenValidityInMilliseconds attribute is initialized
  And A new JWT Filter is initialized with the Token Provider
  And The Security Context's Authentication Token Object is set to null
  And A "wrong" String representation of a token is created
  And A new Mock HTTP Servlet Request is initialized and its Authorization
    header is set with the wrong String Token
  And the Mock HTTP Servlet Request's Request URI is set
  And A new Mock HTTP Servlet Response is initialized
  And A new Mock Filter Chain is initialized
  When The JWT FIlter's doFilter method is called with the Mock HTTP Servlet
    Request, Mock HTTP Servlet Response, and Mock Filter Chain
  Then The Status of the HTTP Servlet Response is 200
  And The Security Context's Authentication property is null

  (*@\textcolor{gray}{Metadata:} @*)
    (*@\textcolor{gray}{ProjectName: loja} @*)
    (*@\textcolor{gray}{FileName: JWTFilterTest.java} @*)
    (*@\textcolor{gray}{TestName: JWTFilterInvalidToken} @*)
    (*@\textcolor{gray}{Codes:} \textcolor{blue}{Token-Authentication, Filter-Chain, multiple-assertion}@*)

\end{lstlisting}
\end{lstfloat}

\vspace{-5pt}

\subsection{Constant Comparative Analysis}
In open coding, concepts are organically extracted from the data rather than assigned from an existing set. Initial codes are often broad, but additional analysis will reveal more specific patterns and concepts. To accommodate this in our analysis we incorporated \textit{constant comparative analysis}~\cite{glaser1967discovery} into the open coding process.

Test methods \& memos were constantly compared to identify reoccurring themes and note variations. New codes were constantly compared against the existing set to merge codes that were too fine-grained (not enough data points) and refine overly-broad codes.

For example, one initial code was \textit{tokens}, which are objects used to store and authenticate user credentials. As more test methods were reviewed this code was refined into 3 new codes representing these actions: \textit{verify token properties}, \textit{token authentication}, and \textit{refresh token}. Another initial code, \textit{verify user properties} was merged with \textit{verify token properties} because further analysis revealed that in all cases, these properties were being retrieved from a token.

\subsection{Selective Coding}
After all the unit tests were annotated, the authors compared their codes to review concepts and themes discovered. During this discussion it was revealed that the author's codes had different focuses: one author focused on \textit{authentication concepts} (\eg{} tokens) and the other focused on properties making tests \textit{smelly or unique}. After reviewing each other's codes, it was decided that each author would then conduct \textit{selective coding}~\cite{glaser1967discovery} on codes for their respective theme, conducting deeper analysis to refine the identified concepts (\cf{} \ding{173} in Figure~\ref{fig:coding-process}).

\subsubsection{Test Cases for Token Authentication\nopunct}
For each code, all relevant unit test methods were analyzed in-depth to capture all unique combinations of test \textit{contexts}, \textit{actions}, \textit{conditions}, \& \textit{expected outcomes}) associated with the code. Constant comparative analysis was applied to modify existing codes to incorporate these combinations. For example, \textit{verify token properties} was refined into four new codes: \textit{input validation during token initialization}, \textit{pre-authentication verification of token properties}, \textit{post-authentication verification of token properties}, and \textit{token validation fails with incorrect input}.

As selective coding progressed, multiple tests sharing the same unique combination of these elements were iteratively merged into natural language \textbf{test cases}. The resulting collection of unique authentication test cases grouped by code form the basis for the study's conclusions about \textit{test cases for authentication}.

\subsubsection{Test Smells\nopunct}
Test cases annotated with smell-based codes were revisited to map these codes to the existing taxonomy of test smells.
A test smell ``refers to a symptom in test code that indicates a deeper problem''~\cite{DBLP:journals/jss/GarousiK18}.
As a starting point for coding and naming the smells, we reviewed the 3 categories of 15 test smells provided by Meszaros~\cite{meszaros2007xunit}.
A more detailed taxonomy from a literature review conducted by Garousi \etal{}~\cite{DBLP:journals/jss/GarousiK18} ultimately served as the reference used to define and identify occurrences of known test smells. However, during the selective coding we intentionally did not limit the search for these known test smells in order to identify new smells and also cover saliences we noticed.
The smells we found are discussed in detail in Section~\ref{sec:smells}.

\subsection{Memo Sorting}
During the selective coding phase, the memos related to authentication concepts were merged to form natural language test cases. These and the test smell memos were then \textit{sorted} by concept to relate test cases to corresponding smells (\cf{} \ding{174} in Figure~\ref{fig:coding-process}).

\subsection{Theoretical Coding}
In the open coding phase, authentication testing concepts were discovered and represented as codes, and \textit{memos} were created to summarize unit tests. Selective coding was then performed to determine the scope and magnitude of each code, resulting in a detailed report of test smells and an organized collection of generalized token authentication test cases. Next, \textit{theoretical coding}~\cite{glaser1967discovery} was performed to connect related codes and construct a taxonomy of token authentication unit tests (\cf{} \ding{175} in Figure~\ref{fig:coding-process}). This would be used to construct the token authentication unit testing guide.

\subsubsection{Test Cases for Token Authentication\nopunct}
During theoretical coding, we integrated and structured our concepts using a hierarchical \textit{coding paradigm}. High-level authentication concepts were structures into \textbf{features} such as \textit{Token Manipulation} and \textit{Login}. Concepts reflecting unique combinations of \textit{context} and \textit{action} related to each feature were structured as \textbf{scenarios}. For example, the \textit{Token Authentication} feature has four scenarios: \textit{token passed to authenticator}, \textit{token passed to authenticating servlet filter via HTTP request}, \textit{application supports multiple authentication types}, and \textit{authentication succeeds}. In this paradigm, test cases within a feature can share a scenario and an \textit{expected outcome}, but have a unique \textit{condition}. 
For instance, 4 test cases in the \textit{token passed to authenticator} scenario have the expected outcome \textit{authentication fails}. However, there are four unique conditions that should cause this outcome: (1) the token value is empty, (2) the token is expired, (3) one or more user details are missing, and (4) one or more user details are incorrect. Each condition could cause a failure or defect if not handled correctly, and all should be verified.This hierarchy was adapted from the concept of \textit{equivalence class partitions} used in test planning~\cite{burnstein2006practical,patton2006software}.

\subsubsection{Test Smells\nopunct}
In selective coding, smell-based codes were mapped to known test smells when appropriate. During theoretical coding, we identified links between the smell concepts and authentication test concepts. For example, it was found that the \textit{obscure test} smell occurred primarily in test cases exercising scenarios for the Token Manipulation feature, specifically \textit{verifying correctness of token properties}. In Section~\ref{sec:smells}, the identified test smells are described and links observed from our dataset are discussed.
%
\section{Test Cases for Token Authentication}
\label{sec:testguide}
At the end of the analysis, we identified \textbf{\numTestCases{}} unique authentication unit test cases, organized by 5 features \& 17 scenarios as described below. For each feature a \textit{flow diagram} visualizes the relationships between the components that make up each test case: a \textbf{scenario} representing the test's \textit{context} and/or \textit{action}, a \textbf{condition} describing the state of an object/property under test during or before the action, and an \textbf{expected outcome}.

\subsection{Token Authentication}\label{sec:token-authentication}
\begin{figure}
\centering
\includegraphics[width=0.47\textwidth]{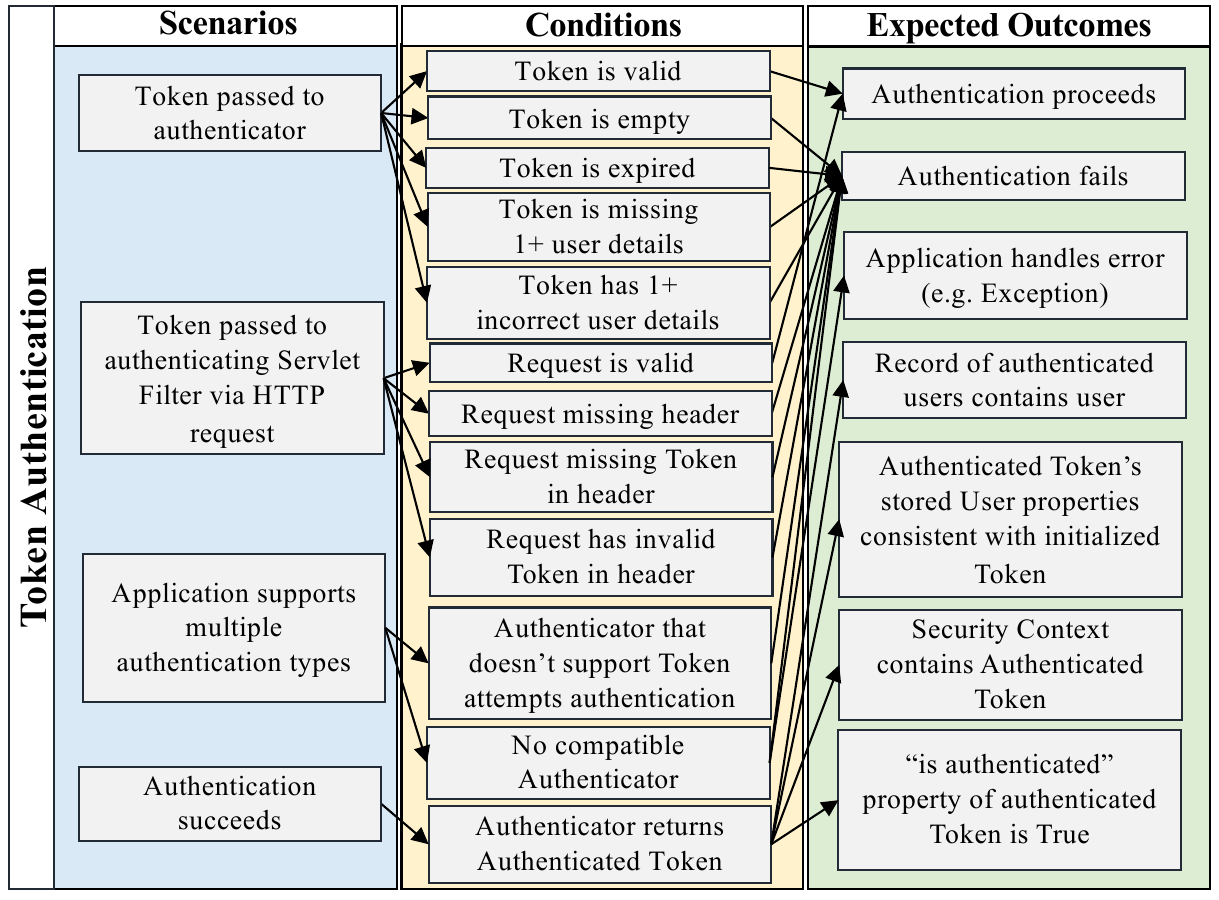}
\vspace{-10pt}
\centering\caption{Test Cases for the Token Authentication Feature}
\label{fig:tokenauth}
\vspace{-10pt}
\end{figure}

Figure~\ref{fig:tokenauth} shows the 4 scenarios and 16 test cases for using tokens to store and authenticate user credentials. A \textbf{token} stores the user's principal (e.g. username) \& credentials (e.g. password) during and after authentication. Test cases for this feature focus on scenarios that occur during authentication of a token that has already been initialized with user data.

The first scenario is \textit{Token passed to authenticator}. Tests for this scenario ensure that authentication fails when the object responsible for authenticating is given tokens containing 4 different types of invalid user data, such as an incorrect password. An additional test checks the `happy path', where authentication proceeds when the authenticator is given a token with valid data. The second scenario is \textit{Token passed to authenticating Serlvet Filter via HTTP request}. Test cases for this scenario verify that when a Filter intercepts and attempts to authenticate incoming HTTP requests, authentication fails when the request is invalid, such as one without a header or with an invalid token. Similar to the first scenario, there is an additional test case ensuring that authentication proceed with a valid request. Next is the \textit{Application supports multiple authentication types} scenario. Many applications support different types of authentication: Username/Password, OpenID, OAuth 2.0, etc. For example, consider many websites allow you to login using your Facebook or Google credentials. The 2 tests for this scenario verify that when an incompatible authenticator attempts authentication or there is no compatible authenticator, then authentication is not successful. For example, a JAAS authentication token should not be successfully authenticated by an OAuth authenticator. The last scenario is \textit{Authentication succeeds}. The 3 tests for this scenario verify the state of the application reflects this: the user should be recognized as an authenticated user, the Security Context should contain the authenticated token, and the user's properties should be stored there correctly.

\subsection{Token Manipulation}\label{sec:token-manipulation}
\begin{figure}
\centering
\includegraphics[width=0.47\textwidth]{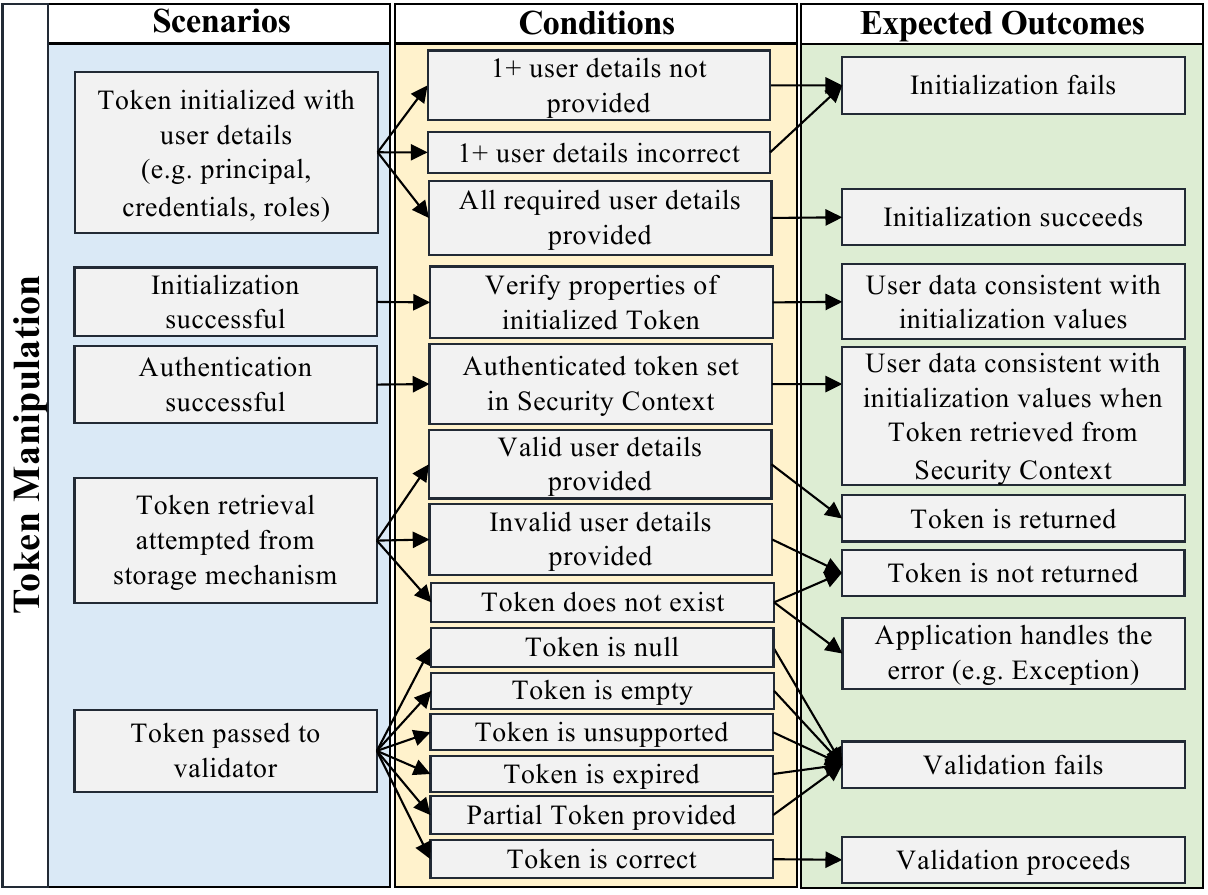}
\vspace{-10pt}
\centering\caption{Test Cases for the Token Manipulation Feature}
\label{fig:tokenmanip}
\vspace{-8pt}
\end{figure}

Figure~\ref{fig:tokenmanip} shows the 5 scenarios and 15 test cases related to \textit{manipulating} tokens used for authentication. This a distinct feature from Token Authentication that focuses on the state of the token and it's stored properties \textit{before} and \textit{after} authentication occurs. Applications should be tested to make sure they handle token creation, access, and validation scenarios with correct \textit{and} incorrect inputs.
The first scenario for this feature is \textit{Token initialized with user details.}  The 3 tests for this scenario check that initialization fails when one or more of these user details (such as username or password) are missing or incorrect, and proceeds when valid details are provided. Next is \textit{Initialization successful}: the test for this scenario ensures that the user data stored in the newly-initialized token are consistent with the provided values when retrieved. Once the token has been created, but before authenticating, the user data properties stored in the token should be checked to ensure it was stored correctly and can be retrieved.  These properties should also be verified after \textit{Successful authentication}: this scenario is distinct from it's companion in the token authentication feature because this test case evaluates consistency of the authenticated token's user data \textit{before} it is stored in the application's \textit{Security Context}, which stores all active security information.
The fourth scenario is \textit{Token retrieval attempted from storage mechanism}. For authentication implementations with storage mechanisms (e.g. OAuth 2.0), token \textit{retrieval} should be tested to ensure that the application validates inputs for token requests. Two test cases for this scenario check that no token is returned for requests with invalid user data or for non-existent tokens. A third test ensures that the application raises an exception for the non-existent case. Finally, a token should be returned when valid user details are provided in a request for an existing token. The fifth scenario is \textit{Token passed to validator} containing five tests; validation should fail when an invalid (null, empty, unsupported, or partially provided) token is given to the validator, and proceed for a valid token.

\subsection{Refresh Token}
\begin{figure}
\centering
\includegraphics[width=0.47\textwidth]{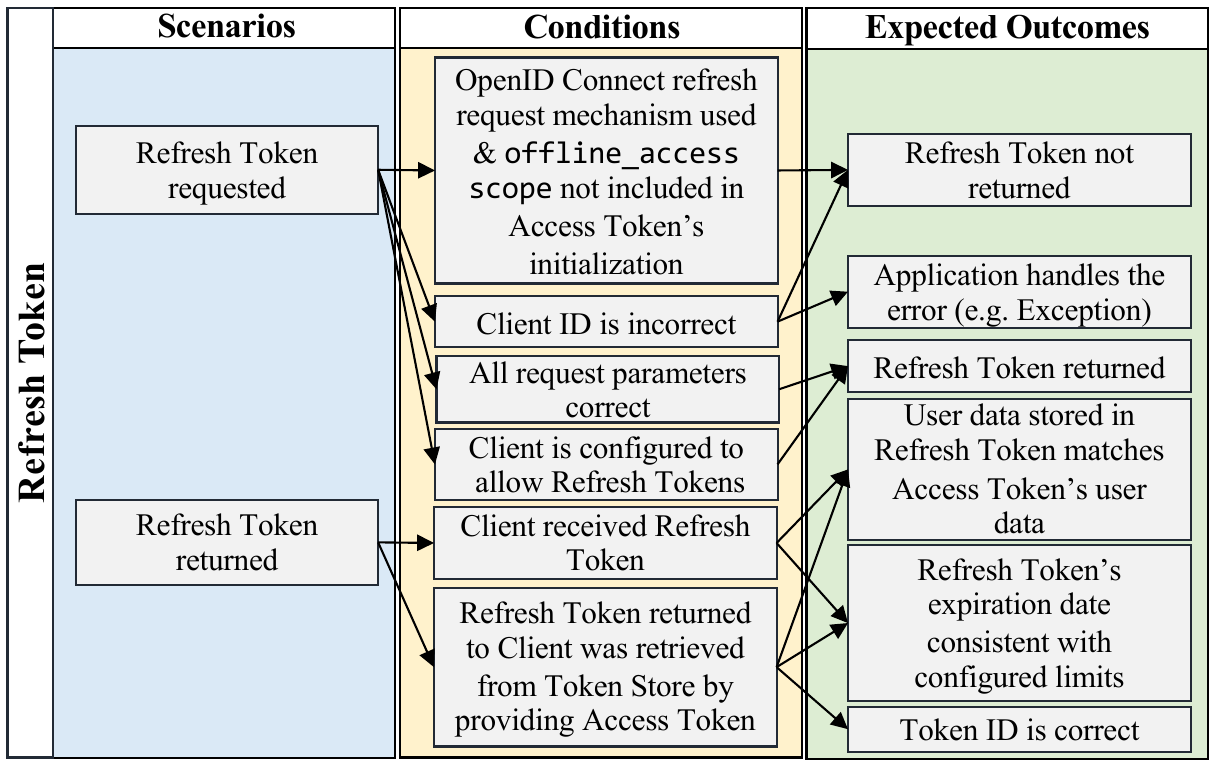}
\vspace{-10pt}
\centering\caption{Test Cases for the Refresh Tokens Feature}
\label{fig:refreshtoken}
\vspace{-20pt}
\end{figure}

Figure~\ref{fig:refreshtoken} shows the 2 scenarios and 10 test cases related to \textit{Refresh} tokens, which are specific to OAuth 2.0 authentication. These tokens are \textit{renewals} of an original ``access" token. This approach enables tokens to have short expiry dates, and generating new tokens does not require frequent re-authenticating. Applications supporting this refresh tokens should be tested to ensure that the properties of new tokens are consistent with the original token and creation/access behaviors only succeed with valid inputs.
The first scenario is \textit{Refresh token requested}. Two test cases for this scenario ensure that a token is not returned when the Client ID is incorrect or (for applications using the OpenID Connect refresh request mechanism) the ``offline access scope'' parameter is not included in the access token's initialization. Another test checks that when request parameters are correct and the client is configured to allow refresh tokens, a refresh token is returned. 5 tests verify that when the client receives the requested refresh token (from a server or Token Store), the user properties are consistent with those in the original token, the new expiry date is correct, and the token ID is correct.

\subsection{Login}\label{sec:login}
\begin{figure}[ht]
\includegraphics[width=0.47\textwidth]{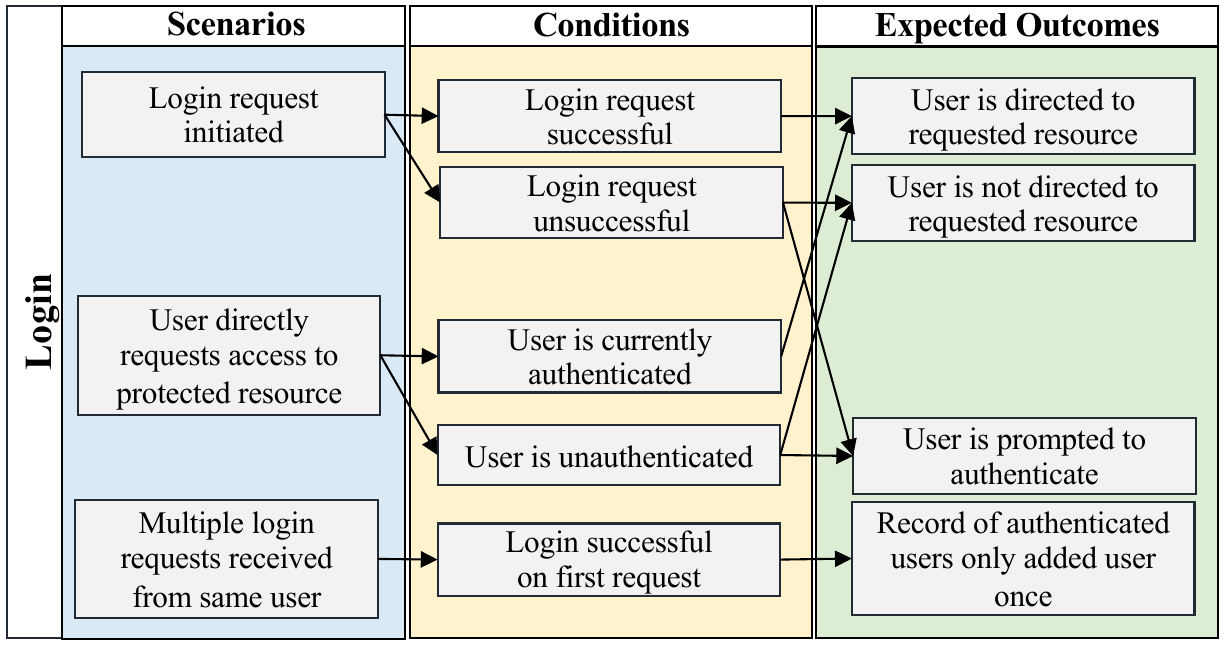}
\vspace{-10pt}
\centering\caption{Test Cases for the Login Feature}
\label{fig:login}
\vspace{-10pt}
\end{figure}

Figure~\ref{fig:login} shows the 3 scenarios and 7 test cases for the Login feature. Login behavior should be tested to ensure that the application appropriately responds to successful \& unsuccessful login form submissions and unconventional request behaviors.
The first scenario, \textit{Login request initiated} is related to form submission and its 3 test cases ensure that users are directed to the requested resource if login is successful and unsuccessful login results in the user being denied access and (re-)prompted to authenticate. The second scenario is \textit{User directly requests access to the protected resource}. For example, consider an attempt to navigate directly to your personal Facebook page instead of the homepage. The 3 tests for this scenario ensure than an authenticated user is granted access and (similar to failed login) unauthenticated users are denied access and prompted to authenticate. Finally, if \textit{ Multiple login requests are received for the same user} (e.g. repeated clicks of the `submit' button), the application's record of authenticated users should not have a duplicate entry for the user.

\subsection{Logout}\label{sec:logout}
\begin{figure}
\centering
\includegraphics[width=0.47\textwidth]{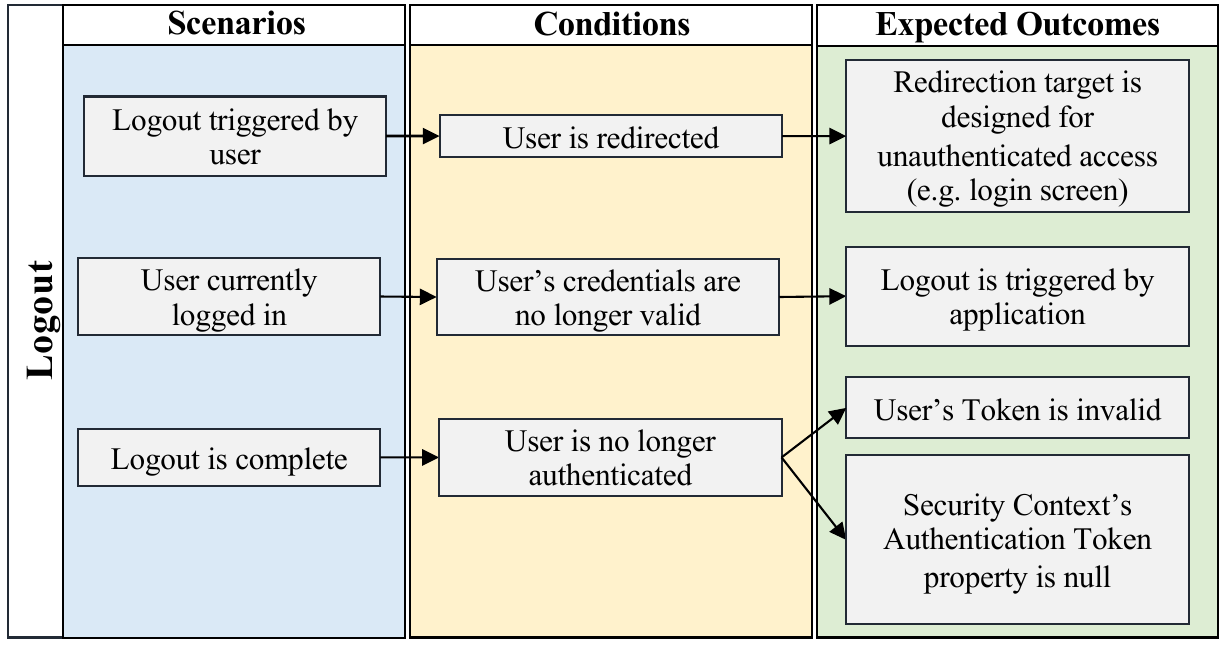}
\vspace{-10pt}
\centering\caption{Test Cases for the Logout Feature}
\label{fig:logout}
\vspace{-10pt}
\end{figure}

Figure~\ref{fig:logout} shows the 3 scenarios and 4 test cases for Logout, the complement to the Login feature. Logout behavior should be tested to ensure that the application state is updated and the User is redirected appropriately.
The first scenario is \textit{Logout triggered by user} (\eg{} hitting the `logout' button) and 1 test case ensures that the user is redirected to a designated location that is `safe' for unauthenticated users, such as a login screen. The second test case is for the \textit{User is currently logged in} scenario. Applications should be expected to automatically trigger a logout if the user's credentials are no longer valid. For example, if you change your Facebook password on a computer, you should be automatically logged out of the mobile app until you re-authenticate with the new password.

The last scenario, \textit{Logout complete} has 2 tests that check application state: the user's token should now be invalid and the Security Context should have a null authentication token.

\subsection{Discussion}
A majority of the identified authentication scenarios in the unit testing guide are concerned with tokens and user credentials. We believe the prevalence of these tests is due to the design of the Spring Security API, which provides \textit{interfaces} for many classes used for the creation and retrieval of user details (\eg{} username, password) or token manipulation with OAuth2. Framework users are required to implement application-specific methods for these interfaces, resulting in the many scenarios from the dataset focusing on these properties.


\section{Analysis of Test Smells}
\label{sec:smells}
Despite plentiful guidelines for developers on how to write high-quality test code, best practices are not always followed in practice resulting in symptoms called \emph{test smells}~\cite{DBLP:journals/infsof/Garousi-YusifogluAC15}.
Contrasting test bugs (faults) in test code that change the intended (expected) behavior, test smells are issues that lower the \textit{quality} of the test, including low fault detection power~\cite{DBLP:conf/icsm/VahabzadehF015}.
In the context of authentication testing, this poses a potential security threat.
This section gives an overview of the test smells we found in the dataset.
Due to space limitation, code examples for the smells are available in supplementary material: \url{https://doi.org/10.5281/zenodo.3722631}

\subsection{Well-Known Test Smells}
In this section we summarize 6 test smells~\cite{meszaros2007xunit,DBLP:journals/jss/GarousiK18} that were repeatedly observed in our dataset and when applicable discuss smells that occurred most often in tests for specific features \& scenarios.

\textbf{Obscure Test} This smell describes difficulties with test comprehension~\cite{meszaros2007xunit,stackoverflow:smells}.
%
It commonly stems from an ``eager" implementation, where the test verifies too much functionality.
For example, instead of providing dedicated test cases, the developers decided to combine similar behaviors (``piggyback'') into one long (``the giant'') test case.
This results in higher maintenance costs, difficulties in understanding the test, and precludes achieving the goal to use \textit{tests as documentation}.
It also may hide many more errors, because during test execution, the first failed assertion aborts the test case.

In our dataset, we observed many obscure tests for the Token Authentication feature (\cf{} Section~\ref{sec:token-authentication}).
Particularly, the same test case may try to authenticate as anonymous user, as guest user, as administrator and as user with a disabled account.
This all happens in sequence instead of writing dedicated test cases for each user type.
In the \texttt{geoserver} project, we found overly long test cases spanning \(>100 \) lines of code, making the test case hard to understand as a whole.
In fact, the developers even placed comments in the test cases to explain what is being tested indicating the developers found the code difficult to understand. However to fix the smell, the test case needs to be restructured and not just commented.

\textbf{Conditional Test Logic} The test case contains code that may or may not be executed~\cite{meszaros2007xunit,koskela:2013}.
For example, we observed several instances of this smell in tests for the \textit{Token passed to authenticating servlet filter via HTTP request} scenario. For example, some tests traversed different configurations in a loop, so the action performed during the test was dependent on the current configuration. This smell complicates reasoning about the actual authentication code and the path taken during test execution; which branch was taken or which iteration failed.

\textbf{Test Code Duplication} In this smell, the same test code is repeated many times~\cite{meszaros2007xunit}. Duplicated code impacts test code maintenance~\cite{gonzalez2017large}, which can be a serious issue: if test logic in on code fragments is modified, all duplicates must be identified \& modified as well.
We observed two manifestations of this: either whole test cases were duplicated (with slight modifications), or code chunks within a test case were repeated.
We found this smell mostly in test cases for authentication protocols, \eg{} SAML or OAuth.
Instead of parameterizing the test cases, the whole test case is duplicated with minor changes, \eg{} testing OAuth with different identity providers like Google or Facebook.

\textbf{Assertion Roulette} In test cases affected by this smell, it is hard to tell which of several assertions failed~\cite{meszaros2007xunit}.
This smell, also called the ``free ride"~\cite{anatasov:2014}, is related to the Obscure Test smell and Listing~\ref{listing:example-test} shows an example. When this test fails during test execution, it is not obvious to tell which assertion failed. Although the number of assertions in Listing~\ref{listing:example-test} is low, the test case is still considered smelly because of missing assertion messages.
This smell was commonly found in tests for the Token Authentication and Manipulation features. A single test would have assertions for multiple user types credentials such as unknown user, normal user, and administrator. Another example is one test for multiple user detail conditions such as different kinds of invalid passwords.

\textbf{Excessive Mocking} This smell, also known as ``mock happy'' refers to a test case which requires many \emph{mocks} to run~\cite{scruggs:2009,trenk:2013}.
The system under test may need a complex environment, specific resources, or depend on 3rd party systems which cannot be provided during test execution. Running tests against a database is a canonical example.
If the database isn't available, all tests will fail even though the system under test might be completely bug-free.
A common solution is to \emph{mock} those potential failures by mimicking the behavior of real objects.
Developers also use mocks when testing authentication, but sometimes struggle with strategy. For example, in StackOverflow question ``Spring Test \& Security: How to mock authentication?''\footnote{\url{https://bit.ly/2ZLFV4U}} the user wants to write unit tests to verify that URLs of the developed web service are properly secured.
Spring Security has a built-in solution for using mocks in tests, which was also picked as answer for the referenced question. However, these built-in solutions were rarely applied in our dataset; instead developers came up with individual ones leveraging 3rd party mocking frameworks.
Further, in some test cases the amount of mocked objects and behaviors is overwhelming, and it is hard to understand what is actually ``real'' and being tested.
The many mocks indicate many dependencies, making the test fragile and resistant to change.

\textbf{Issues in Exception Handling} This ``silent catcher'' smell refers to a test case that succeeds even when an exception is thrown~\cite{DBLP:journals/jss/GarousiK18}.
Mostly, the exception is ignored along with its type, which represents specific semantics.
This may lead to false results, \eg{} when testing the authentication protocol.
For example, consider a test whose intended behavior is to check whether an exception is thrown when invalid credentials are provided. Without detailed handling of the exception, the test passes even if the exception that actually occurs is one that the developer did not expect, such as if the database storing the users is not available.

\subsection{Novel Test Smells}
While analyzing the test cases for smells we also noticed interesting tests containing previously unreported test smells.

\textbf{Testing the Authentication Framework}
In project \texttt{LuckyPets}, we found a test case that validates the correct behavior of the \emph{SHA-1} cryptographic hash implementation provided by the Spring framework.
The test literally compares the generated message digest with a string constant to see if both are identical.
Since the hash implementation is part of a 3rd party library, the library maintainers are responsible to properly test the code and not the developer.

\textbf{Mocking a Mocking Framework} Similarly, we found multiple occurrences where test cases tested the applied mocking framework.
There, the test author assembles a specific object, \eg{} a user, and instructs the mock library to return this object when the function under test is called.
After performing this call, it is asserted whether the returned object is identical to the one the mock library was instructed to return.
Thus the correct behavior of the mock framework is tested and not that of the function.

\subsection{Discussion}
The above smells are not specific to authentication testing.
However, some smells occurred more often in tests for specific authentication components.

The obscure test smells was often found when testing \textbf{user credentials}.
There, a few long complex test cases test different credentials combinations such as default user and password, no password, admin user, disabled user, and of course a successful login.
As a result, a failing test case does not directly provide feedback which credential combination failed.
Spring Security uses filter chains to intercept requests, detect (absences of) authentication, redirect to authentication entry points, and eventually let the request hit the servlet or abort with a security exception.
Similarly to user credentials, different setups, \ie{} types and orderings, of filter chains are often tested in the same test case and thus render it obscure.

In the unit tests for \textbf{OAuth}, we often observed \textit{test code duplication}.
In the respective code, only the specific endpoint changes, \eg{} Google or Facebook as identity providers. Duplicating code instead of parameterizing tests increases the maintenance costs.

When testing \textbf{authentication responses}, a common smell we found is \textit{conditional test logic}.
Loops and if/else chains are utilized to generate multiple authentication requests and test the responses (see Sections~\ref{sec:login} and \ref{sec:logout}).
This approach exhibits the conditional logic smell, which hides the actual code paths taken during test executions.
A cleaner solution would be to implement separate test cases with linear code paths.
However, testing responses usually involves mocking, such as the requests and the utilized filter chain.
In this respect, we also noticed the excessive mocking smell, where a multitude of objects are mocked.
Once all these mock objects are created in the test case, they are used multiple times presumably to avoid code duplication.
The same solution could also be achieved by refactoring the common mock assembly steps in setup routines which are called at the start of individual test cases.
\section{Practitioner Perspectives}\label{sec:interviews}
To evaluate the usefulness and accuracy of the testing guide, we sought feedback from practitioners. We corresponded with 4 practitioners from two different groups: \textit{unit test authors}, who wrote unit tests we analyzed in this study and \textit{external practitioners}, who did not contribute to our data set but have experience with implementing authentication in Java.

\subsection{Unit Test Authors}
We identified 11 authors of select test files in our dataset from different projects and reached out to them for a remote interview. Two developers agreed to participate, each a senior developer but with different experience with implementing authentication in Java (see Table~\ref{tab:test-authors}). The goal of these interviews was to discuss their rationale when writing the tests, their general testing strategy, and perspectives on unit testing for security features.
\begin{table}
\centering
\renewcommand{\arraystretch}{0.8}
\caption{Unit Test Authors Interviewed}
\label{tab:test-authors}
\vspace{-10pt}
\begin{tabular}{@{}lp{2.3cm}p{4.6cm}@{}}
\toprule
Test & Background & Experience \\
Author & & \\ \midrule
A-1 & Senior Developer, Co-Founder of \textit{Major OSS Project} & 20 years developing and testing Java \textit{inc.} Spring, about 5 years using Spring \textit{Security} \\
A-2 & Software Developer & 18 years of Java development \& unit testing, little authentication \\ \bottomrule
\end{tabular}
\vspace{-10pt}
\end{table}

\subsubsection{Leading Interview Questions\nopunct}
We developed 5 questions to guide discussion but allow open-ended responses from participants:

\noindent\textbf{Q1}: \textit{Could you describe your technical background?}
This question established the experience of the practitioner. Follow-up questions were asked when needed to determine experience with \textit{Java}, \textit{implementing authentication}, and \textit{unit testing}.

\noindent\textbf{Q2}: \textit{How would you describe your involvement in writing and testing authentication in this project?} We encouraged the developers to discuss the analyzed test cases along with any other authentication-related contributions they made.

\noindent\textbf{Q3}: \textit{How did/do you determine which test cases to write?} This question tells us if they used a metric (e.g. coverage), test guides, or other resources when planning test cases.

\noindent\textbf{Q4}: \textit{Would a unit testing guide for authentication would be helpful?} After asking this question, we asked them if they thought select test cases from our guide, related to the scenarios they tested, would have been useful to them when writing the tests.

\subsubsection{Results\nopunct}
Each developer's background and experience (\textit{Q1}) are summarized in Table~\ref{tab:test-authors}. Regarding \textit{Q2}, Developer A-1 wrote many of their project's initial authentication unit tests. These tests are actually part of the project itself, so they become part of the test suite of any application that uses it. Developer A-1 no longer works on authentication, but their related unit tests remain in use, mostly unchanged except versioning updates. Developer A-2 had 18 years of Java experience but limited experience implementing/testing application security features. About \textit{Q2} he mentioned:
\begin{quote}
\textit{``Q2: Every now and then, in my job the security related features} [such as] \textit{Kerberos, Authentication encryption, and etc.) come on but I'm not an expert in them. Someone has to implement and test them.''}
\normalsize
\end{quote}

\noindent When asked how they decided what unit tests to write, A-1 said:
\begin{quote}
\textit{``Q3: We start our testing effort by increasing coverage, but when we got many users, bugs arrived and then we added more test. These are mostly tests we didn't catch by coverage. Things like Token Provider, JWT filter} (JSON Web Token)\textit{, and security filter were complex, they clearly needed test. Those are the type of things you don't want to leave untested.''}
\normalsize
\end{quote}

\noindent In this regard, A-2 mentioned:
\begin{quote}
\textit{``Q3: [I] did not have an authentication-based perspective, it was much more like getting full coverage''}
\normalsize
\end{quote}


\noindent Finally, when asked if a test guide would have helped them when they wrote the tests (\textit{Q4}), A-1 said:
\begin{quote}
\textit{``Q4: This guide is an improvement, there is no other guide today, so this is definitely useful. Specifying which test is more suitable for unit test and which one is more suitable for integration testing is useful. People will mess this up.''}
\normalsize
\end{quote}

\noindent A-2 said the following statement about usefulness of the guide:
\begin{quote}
\textit{``Q4: At the time, we wanted something that works ASAP, so in the beginning the guide would have not been helpful. However, when we needed to improve our authentication, such through guide on how to test authentication would have been useful.''}
\normalsize
\end{quote}


\subsection{External Reviewers}
We contacted 4 practitioners (from corporations in North America) who had not worked on the systems we studied but considered themselves experienced with implementing authentication in Java. Two of these developers agreed to review our testing guide and provide feedback. Table~\ref{tab:reviewers} summarizes the background and related experience for each reviewer.
\begin{table}
\centering
\renewcommand{\arraystretch}{0.8}
\caption{Practitioner Reviewers}\label{tab:reviewers}
\vspace{-10pt}
\begin{tabular}{@{}lp{2.1cm}p{4.1cm}@{}}
\toprule
Practitioner & Background & Experience \\ \midrule
P-1 & Senior Developer, Architect, SE Researcher & 13--14 years developing and testing Java applications, 3--4 years implementing authentication\\
P-2 & Software Developer & 5 years using Java, implemented authentication about 5 times \\
\bottomrule
\end{tabular}
\vspace{-10pt}
\end{table}

\subsubsection{Feedback Guidelines\nopunct}
Correspondence between authors and the practitioners was conducted through email. The initial correspondence included a brief description of the project adapted from the abstract of this paper. Upon agreement to participate, each practitioner was sent a PDF copy of the testing guide, which did not include Figures~\ref{fig:tokenauth}-~\ref{fig:logout}, and the direction that we would \textit{``appreciate [their] honest opinion, based on [their] experiences''} of the test guide's \textbf{format} (\textit{Is the guide easy to understand/read?}), \textbf{contents} (\textit{Are any important test cases missing, or are any included test cases invalid/unnecessary?}), and \textbf{overall usefulness} (\textit{Would this help you test an authentication implementation, or evaluate coverage from a security perspective?}). We also encouraged reviewers to include `any other comments' they might have.

\subsubsection{Results\nopunct}
Neither practitioner found the guide's \textbf{format} difficult to understand, but P-1 suggested to present test cases as \textit{``\(\ldots \) kind of scenario or a flow \(\ldots \)''} that started with \textit{``all possible scenarios''}. Based on this feedback, we included Figures~\ref{fig:tokenauth}-~\ref{fig:logout} in the final version of the test guide. P-1 also requested \textit{``more description and more details, some terms need to be defined.''}, which has been addressed. P-2 suggested minor changes to the phrasing and grammar of a few tests cases. Neither practitioner thought any test cases (\textbf{contents}) were invalid. In fact, P-1 highlighted the guide's wide applicability: \textit{``I think you covered most of the test cases that come in my mind regardless what authentication technique is used.''} However, P-1 also suggested a \textit{missing} test case: for the \textit{logout is complete} scenario, the token store should be checked to verify the token was removed. This was included in the final version of the test guide, available in our supplementary data package.
Both practitioners agreed that overall, the test guide was \textbf{useful} and helpful. P-1 said \textit{``I would say 70\% I find this guide useful''}, and reiterated their content and formatting suggestions.

\subsection{Summary of Practitioner Perspectives}
Overall, practitioners consider the token authentication unit testing guide to be useful, and none of the included test cases were marked as invalid. The guide's narrow scope and some missing test cases were the only significant drawbacks brought to our attention. Both unit test authors, who are on opposite ends of the \textit{authentication experience spectrum}, felt the guide was useful. Their main concern was, in practice, developers plan testing from the \textit{coverage} perspective. However, A-2 felt they would benefit from using the guide for retrospective or later-stage testing. We incorporated most of the minor formatting suggestions and missing test cases into the final version of the test guide. The coverage comments suggest the guide could be applied towards developing a technique to calculate a \textit{security coverage} metric, which we will explore in future work.

\section{Literature Review}
\label{sec:related}
Post-analysis, we conducted a literature review to examine whether our findings are supported by existing literature or add new ideas to the knowledge base. We organize the related works into 2 groups: \textit{testing authentication}, existing authentication testing resources and \textit{test smells}, existing work on smelly test code. The IEEE Explore, ACM Direct, Springer Link, and ScienceDirect research repositories were manually searched to find full papers. For the first group the search terms were combinations of \textit{`unit test', `test case' `security',} and \textit{`authentication'}. For the second group, the terms were \textit{`code smells'} and \textit{`test smells'}.

\subsection{Testing Authentication}
Numerous resources exist which focus on security testing ~\cite{wysopal2006art,addressingissues,buildingsecuresoftware,stallings2012computer,potter2004software,potter2004software,DEVRIES200711,de2006security}. Most notably the \emph{Open Web Application Security Project (OWASP)} published a guide~\cite{meucci2013owasp} describing techniques for testing web service security issues, and assembled a cheat sheet~\cite{owaspAuthenticationCheatSheet} focusing on specific security topics including authentication. This guide contains more test cases than our work, and includes session management, but the test cases were not applicable from the unit testing perspective. Additionally, a search of major research publications revealed no studies examining authentication unit testing. Therefore, we have \textit{extended} the existing knowledge base with a testing guide that explicitly describes token authentication test cases that can be written as \textit{unit tests}. 
\subsection{Test Smells}
The term ``code smell'' was introduced by K. Beck and M. Fowler and served as motivation for their book about code refactoring~\cite{fowler2018refactoring}. Later researchers specifically looked into smells unique to test code~\cite{van2001refactoring}. These ``test smells'' refer to any symptoms in test code that indicate a deeper problem~\cite{DBLP:journals/jss/GarousiK18}: poorly designed tests and their presence may negatively affect test suites and production code (maintainability and functionality), or reductions in one or more test quality criteria. Meszaros~\cite{meszaros2007xunit} presents 15 test smells organized in three categories: code smells, behaviors smells, and project smells.
A recent literature survey by Garousi and K{\"u}{\c c}{\"u}k~\cite{DBLP:journals/jss/GarousiK18} extracted a large catalog of published test smells. Additionally, the work describes negative consequences of the smells and ultimately how to deal with them. We used this catalog and the one of Meszaros~\cite{meszaros2007xunit} to identify and name the smells. Developing high-quality test code is a non-trivial task. To assist developers, many guidelines have been proposed~\cite{DBLP:journals/jss/GarousiK18}. Garousi \etal{}~\cite{DBLP:journals/infsof/Garousi-YusifogluAC15,DBLP:journals/software/GarousiF16} introduced the term \emph{Software TestCode Engineering (STCE)}, which refers to a set of practices and methods to systematically develop \& maintain high-quality test code. Thus, we extend the existing body of test smell knowledge by adding a security perspective, highlighting smells associated with classes of authentication tests.

%
\section{Threats to validity\label{sec:threats}}
We identified potential threats to the validity~\cite{wohlin2012experimentation} of our study and have mitigated them as follows.

\emph{Construct Validity.}
In our study we analyzed test cases to identify common test cases and test smells for unit testing token authentication. This was done by manually coding each test method.
Here the threat is whether the resulting codes can be accurately used for our identification task. To mitigate this threat, we followed the key principles of grounded theory and consistently reviewed the process to avoid deviations. Further, the resulting codes were peer reviewed to ensure consistency.

\emph{Internal Validity.}
The main threat to the internal validity of the research is the manual analysis of unit test methods.
This includes assigning codes to each test case using grounded theory, and identifying test patterns and test smells; manual analysis can be prone to bias and errors.
The coding task was performed by two researchers with multi-year experience of the java programming language, and unit testing.
Emerging codes were constantly compared among existing ones to observe commonalities and differences.
Feedback and interviews with 2 practitioners and 2 authors of tests from our data set were conducted to evaluate whether our findings are valuable for industry. The goal was to show the study's \textit{practical} significance and not \textit{statistical significance}. The practitioners found the results valuable, and important in practice.

\emph{External Validity.}
For our study, we solely focused on open-source projects, since those were the only available projects to us that provided all the necessary information to conduct this study. A potential threat to external validity arises when we want to generalize our findings to a wider population that includes commercial developments. While open-source tools primarily focus on automated unit tests for software quality assurance, the of closed-source projects applied a combination of automated unit tests and manual user acceptance tests. To mitigate generalization issues, the findings extracted from this data are represented in a generic format that is applicable to other security frameworks.
As argued by Wieringa et.al~\cite{Wieringa2014}, describing the context of studied cases as we did allows us to ``generalize`` our findings by analogy; i.e., findings may apply to other security frameworks that provide an API for implementing token based authentication, token manipulation, refresh tokens, and user login/logout. To verify that the derived test cases are generalizable to other \textit{Java} web security frameworks, documentation was reviewed for JAAS and Apache Shiro. Only framework-agnostic objects (e.g. Security Context, Servlet Filter) are identified by name in test cases. Further, including other languages or frameworks would have minimal impact on the final results, as token authentication follows a generic flow~\cite{ScotchIO} and the APIs use language-agnostic protocols such as OAuth2 (RFC 6749) and OpenID Connect.
%
\section{Conclusion}\label{sec:conclusion}
This work analyzed authentication unit testing in Java using the Spring Security framework.
Our semi-automated approach created a detailed taxonomy of \numTestCases{} unique test cases as well as common test smells to avoid.
Our source for these recommendations were over 400 JUnit test cases extracted from 53 open source Java projects from Github.
We curated our dataset by automatically \textit{mining} candidate projects and using static analysis to detect JUnit test files which imported packages related to authentication from the Spring Security API.\@
Next, we used a grounded theory approach to manually identify concepts in the collected data.
This involved a four stage coding process resulting in the test guide and supplementary analysis of smells. For each test case, a memo containing \emph{context}, \emph{action}, \textit{condition} and \emph{expected outcome} was written.
The memos served as a foundation for creating the test catalog and test smell analysis.
We conducted two interviews with authors of unit tests from our data set and two other selected practitioners to verify that these findings are useful. The developers confirmed the importance, quality, and usefulness of the test guide, and suggested a few test cases that were not discovered in the data.
We supplement this paper by providing the mined dataset of authentication unit test files, the memos and codes used in analysis, and the complete testing guide: ~\url{https://doi.org/10.5281/zenodo.3722631}.
We hope that these recommendations are used by developers to determine which test cases to write for their authentication implementations, or ensure what they've written is complete.

\section*{Acknowledgments}
\small{This work was partially supported by: NSF award CNS:1816845, DFG grant MA 5030/3-1, DLR grants D/943/67258261, D/943/67262000, and BMBF grant 01IS18074E.}

\balance{}

\bibliographystyle{ACM-Reference-Format}
\bibliography{acmart}

\end{document}